\documentclass[12pt]{iopart}

\usepackage{iopams}
\usepackage{graphicx}
\usepackage{color}
\usepackage[utf8]{inputenc}

\begin{document}

\title[Potential Energy Curves and GOS for Doubly Excited States of H$_2$ Molecule]{Potential Energy Curves and Generalized Oscillator Strength for Doubly Excited States of Hydrogen Molecule}

\author{L. O. Santos$^{1}$, A. B. Rocha$^{2}$, R. F. Nascimento$^{3,4}$, N. V. de Castro Faria$^{1}$ and \\ Ginette Jalbert$^{1}$}
\address{$^1$Instituto de F\'\i sica, UFRJ, Cx. Postal 68528, Rio de Janeiro, RJ 21941-972, Brazil \\
$^{2}$Instituto de Química, Universidade Federal do Rio de Janeiro, Rio de Janeiro 21941-909\\
$^{3}$CEFET/RJ, UnED Petr\'{o}polis, RJ 25620-003, Brazil\\
$^{4}$Department of Physics, Stockholm University, AlbaNova, SE-106 91 Stockholm, Sweden
}
\ead{leonardosantos@if.ufrj.br}

\begin{abstract}

{In this paper we report calculations of potential energy curves in the $1.2 a.u.\le R\le100 a.u.$ range at Multireference Configuration Interaction (MRCI) level for doubly excited states of the H$_2$ molecule. We have focused on the $Q_2$ states which lie between the second and third ionization thresholds of H$_2^+$ molecular ion, i.e., $^2\Sigma_u^+$ state in which lie the  H(2l) + H(2l') dissociation channels. The MRCI approach allowed us to successfully identify for the first time the molecular state which dissociates into hydrogen atoms at 2s state. Further, Generalized Oscillator Strength as a function of transferred momentum for three doubly excited states is also presented.} 

(Some figures in this article are in colour only in the electronic version)
\end{abstract}

\pacs{34.10.+x, 31.15.vn, 31.50.Df, 34.80.Gs, 34.80.Ht, 34.50.Gb}
\submitto{\JPB}
\maketitle

\section{Introduction}

One of the first evidences of the autoionized doubly excited states of H$_2$ was obtained in the experiment of Crowe and McConkey in 1973 \cite{Crowe} in which electrons collided with H$_2$ at impact energies of 29, 33, 50, 100 and 399 eV. The data revealed the existence of new unknown states besides the ground state $^2\Sigma_g^+$ and the first excited state $^2\Pi_u$ of H$_2^+$. The first theoretical calculations of the states were made by Bottcher and Docken in 1974 \cite{Bottcher}, who calculated the energy position and widths for dissociative  $Q_1$ $^1\Pi_u$ and $Q_2$ $^1\Pi_u$ states for $1a.u.\le R\le 10a.u.$, using the Feshbach projection-operators \cite{Feshbach}. More recently Sánchez and Martín \cite{Martin} have done systematic calculations in which they estimated tens of doubly excited states symmetries and their autoionization widths for $^{1,3}\Sigma_{g,u}^+$, $^{1,3}\Pi_{g,u}$ and $^{1,3}\Delta_{g,u}$, in the range of $0\le R\le 5 a.u.$. They employed the Feshbach projection-operator method and a $L^2$ representation of the non resonant continuum, with B-spline type functions. In what concerns the dissociation into neutral atoms, perhaps the most relevant theoretical work in the recent years was done by Dalgarno and collaborators \cite{Dalgarno} who calculated the doubly excited H$_2$ states converging to H(n=2) + H(n'=2) for all the internuclear distances. Using different techniques to solve for each internuclear distance region and taking special care of intermediate distances in order to connect the different regimes in which calculations were done, they were able to calculate potential energy curves from 3 a.u. to 200 a.u. Recently our experimental group \cite{Grupo1} observed the pair H(2s) + H(2s) by measuring neutral fragments in coincidence after H$_2$ dissociation induced by electron impact, confirming for the first time the existence of such states. The purpose of the present work is to shed light on the description of the doubly excited states that may dissociate onto a pair H(2s) + H(2s), since this dissociation channel, as discussed in \cite{Bohm}, may be used to probe the spin coherence between the two hydrogen atoms in future applications. For this, we have used an approach that mixes the standard CI \cite{CI} method with MCSCF \cite{MCSCF}, known as MRCI \cite{Botch} in the treatment of two-electron diatomic molecules allowing us to rise accurate potential curves at every internuclear distance. In addition, since we intend to work with our experimental group \cite{Grupo1}, is desirable to provide it with cross section calculations within the First Born Approximation, that would be helpful in the experimental apparatus adjustment. For this purpose, we begin by presenting Generalized Oscillator Strength calculation for some doubly excited states. Finally, it is important to note that the main feature of our calculations is to provide for the first time an unambiguous characterization of the electronic state that goes, in the dissociation limit, to H(2s) + H(2s) fragments.

We start describing the theoretical methods used for the electronic state energies and generalized oscillator strength. Results for three molecular regimes, Franck-Condon (FC) (1.2a.u. - 3.0a.u.), intermediate (3a.u. - 12a.u.) and asymptotic range $(\geq 80a.u.)$. We also calculate generalized oscillator strength for those states. At last, conclusion and perspectives are pointed out.

\section{Theoretical method}

Atomic units (e=$m_e$=$\hbar$=1) are used throughout, except where indicated. The origin of the electronic coordinates is located at the center of mass of the molecule and has its z-axis lying along the internuclear axis. 
\subsection{Electronics States}
In this study we implement the Huzinaga basis-set \cite{Dunning}, similar to that used by Borges {\it et al} \cite{Itamar}, which consists of cartesian gaussian-type (12s, 6p, 3d, 1f)/[9s, 6p, 3d, 1f] and provides a total of 110 molecular orbitals: $29\sigma_g$, $29\sigma_u$, $22\pi_g$, $22\pi_u$, $4\delta$ and $4\phi$. 
The electronic states are obtained within the Born-Oppenheimer approximation, making use of a hybrid technique, which is a mixture of the Multi-Configurational Self-Consistent Field (MCSCF) \cite{MCSCF} and the Configuration Interaction (CI) \cite{CI} methods (also called multi-reference CI (MRCI) \cite{Botch}). We perform a state averaged MCSCF (SA-MCSCF) over a set of 20 states, all of the $\Sigma_g^+$ symmetry. These states are used here for two reasons: first, they allow us to locate the position in energy of the states of interest, especially those of the $\Sigma_g^+$ symmetry which dissociate in H(2s) + H(2s); second, we are able to optimize the orbitals of interest which generate the $Q_2$ states which have $\lambda=\vert m_z\vert$ = 2, where $m_z$ is the projection of the electronic angular momentum $L_z$ along the internuclear axis. For this purpose, we have used an active space composed of 12 molecular orbitals, $n\sigma_g$ (n = 1-3), $n\sigma_u$ (n = 1-3), $n\pi_g$ (n = 1-2), $n\pi_u$ (n = 1-2) which comprise the main configurations of the electronic states from the Franck-Condon to the dissociative region. Since we are interested in the $Q_2$ states, we have excluded the $1\sigma_g$ and $1\sigma_u$ orbitals of the configurations at the CI step. However, in order to generate both $\mathbb{P}$($\vert 1s\sigma_g\rangle$ + $\vert 1s\sigma_u\rangle$) and $\mathbb{Q}$ = ($1-\mathbb{P}$) \cite{Feshbach} precise spaces, such orbitals are still calculated at the MCSCF step. Hence, potential energy curves of all symmetries at the ($1a.u.\le R\le 100a.u.$) range have been obtained. 

\subsection{Generalized Oscillator Strength}
In this paper we calculate the Generalized Oscillator Strength (GOS) for some doubly excited states in order to evaluate the cross section of excitation of the H$_2$ molecule, in the vertical approximation \cite{Rocha}. In this work, we perform the electronic part of the GOS, which is given by the following expression: \\
\begin{eqnarray}
f(K,E)_{0\rightarrow n}=\frac{2\Delta E}{K^2}\frac{g_n}{4\pi}\int d\Omega\vert\epsilon_{0n}(K,R_0,\Omega)\vert^2
\end{eqnarray}
where $\Delta E$ is the transition energy, K is the module of transferred momentum, $g_n$ is the degeneracy of the excited electronic state and $R_0$, the distance of equilibrium. The integral in $d\Omega$ is performed over all possible orientations of the molecule with respect to {\bf K}. The amplitude of scattering within the First Born approximation (FBA) for a molecule H$_2$ is given by: \\
\begin{eqnarray}
\epsilon_{0n}(K,R_0,\Omega)= -\int d{\bf r}_1{\bf r}_2\psi_n(r_1,r_2;R_0)[\sum_{i=1}^2exp(i{\bf K.r}_i)]\psi_0(r_1,r_2;R_0)
\end{eqnarray}
In order to calculate the generalized oscillator strength, we have implemented the quadrature method of Gauss-Hermite to the electronic coordinates integrals, which works fine in a set of Cartesian Gaussian basis. The electronic states were obtained as previously mentioned, i.e., MRCI approach. One should note the important fact that, in a single determinant approach, the ground state differs from doubly excited states for more than one excitation which, for a one electron operator, would result in a identically zero transition value (Condon-Slater rule \cite{Szabo}). However, our technique takes into account highly correlated wavefunctions, in such a way that the states are not described by only one Slater determinant, but a combination of them (CI), resulting in a calculation of transition matrices of two excitations in only one step.


\section{ Results and Discussion}
\label{sec:fast}

Here we present our results for the potential curves of $Q_2$ states from the Frank-Condon region ($1.2a.u.\le R\le3.0.a.u.$) to the dissociative region ($\approx 100a.u.$) as the Generalized Oscillator Strength, within the vertical approximation, using the MRCI \cite{Botch}. 

\subsection{Franck-Condon range}

\begin{figure}[htbp]
\centering
\includegraphics[width=6.3in,height=6.3in,keepaspectratio]{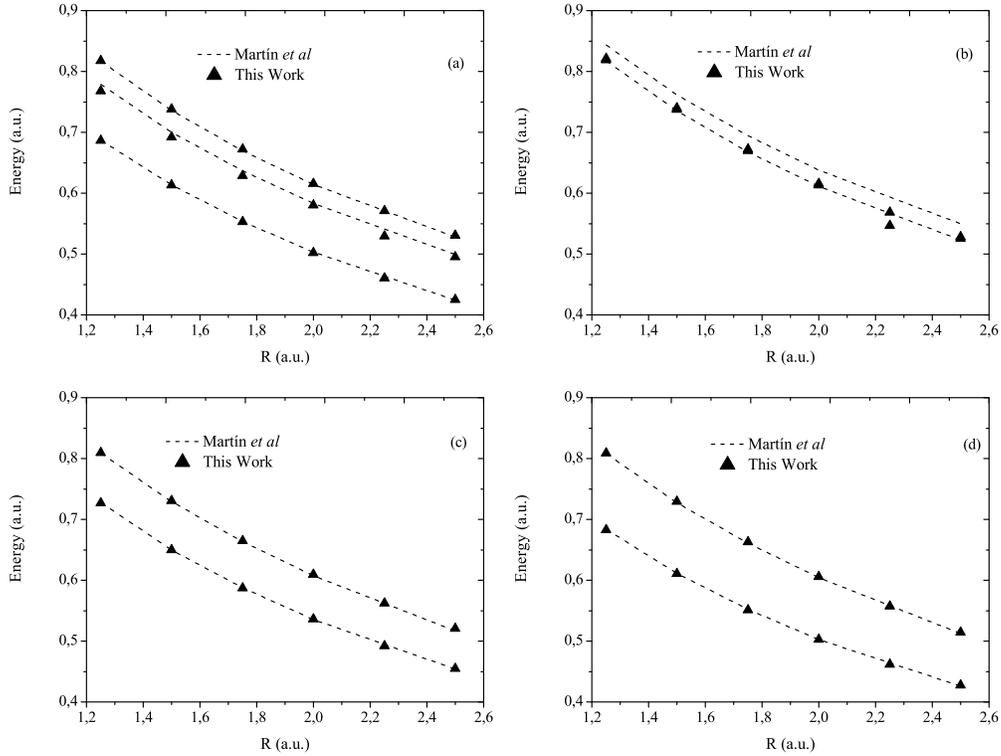}
\caption{Potential Energy Curves as a function of the internuclear distance. Dashed lines, Sanchez and Martín \cite{Martin}; triangles, our results: (a) the three first $Q_2$ $^1\Sigma_g^+$ symmetries; (b) two first $Q_2$ $^1\Sigma_u^+$; (c) $Q_2$ $^1\Pi_u$; (d) $Q_2$ $^1\Pi_g$}
\label{fig1}
\end{figure}

\begin{figure}[htbp]
\centering
\includegraphics[width=6.3in,height=6.3in,keepaspectratio]{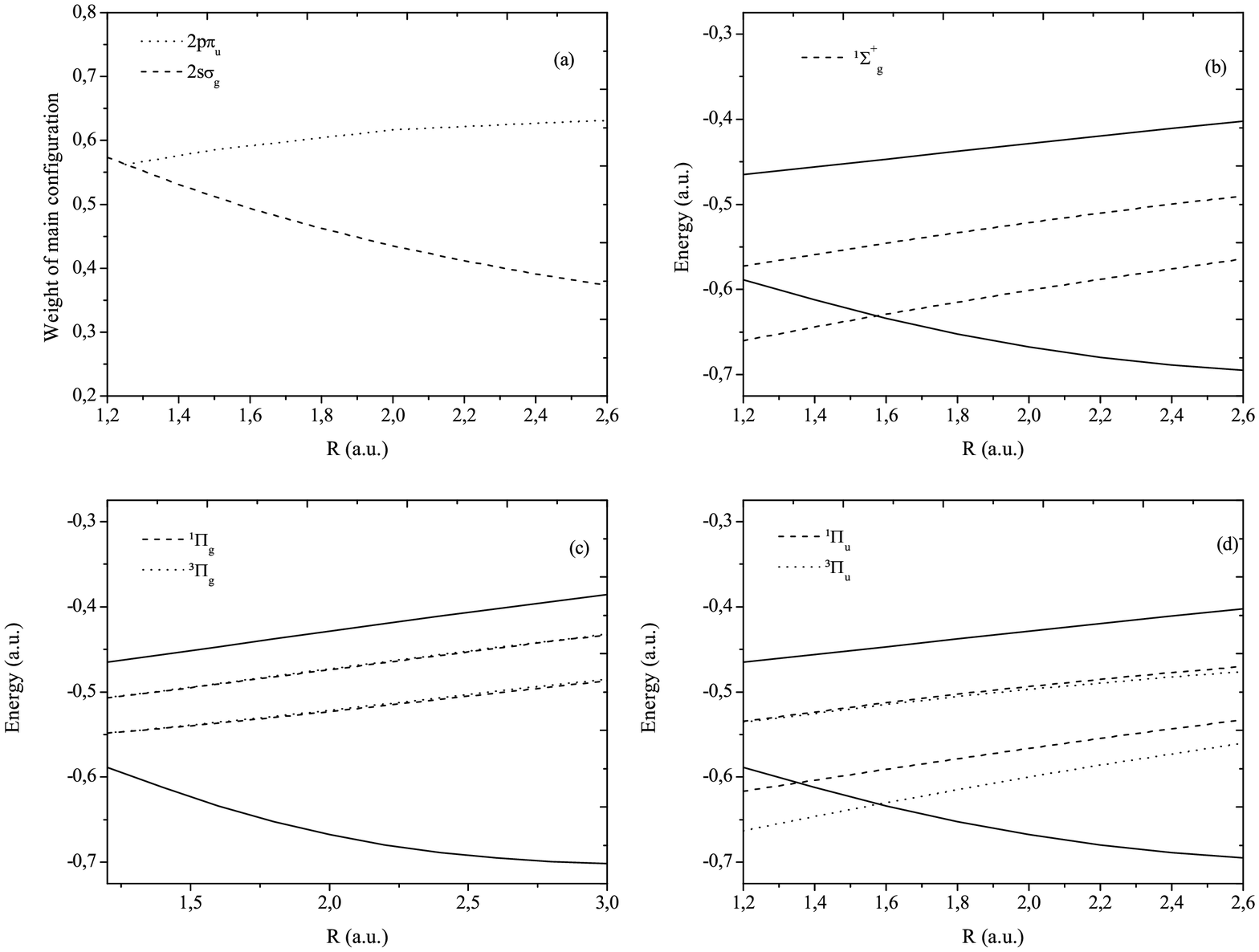}
\caption{(a) Weight of the main configurations of $1Q_2$ $^1\Sigma_g^+$ as a function of internuclear distance: in dotted lines ($2p\pi_u$;$n\lambda\pi_u$); dashed lines correspond to the ($2s\sigma_g$;$n\lambda\sigma_g$). In (b), (c) and (d), solid lines are the $^2\Sigma_u^+$ and $^2\Pi_u$ H$_2^+$ ion energy, dotted and dashed lines are the potential energy curves as a function of the internuclear distance of some electronic doubly excited states ignoring the nuclear interaction}
\label{fig2}
\end{figure}

The FC region is extremely important, especially in collisional processes, since that is the region where the electronic transitions take place. From the theoretical point of view, a reliable calculation of excitation cross sections requires that the wavefunctions are quite accurate in such region. In order to test the quality of our wavefunction, we have calculated the dipole moment of the H$_2$ molecule and show in Table \ref{table1} the results for the excited states $B^1\Sigma_u^+$ and $B'^1\Sigma_u^+$, respectively. Note that the results present an excellent agreement with the exact value calculated by Wolniewicz {\it et al} \cite{Wolniewicz}, with a maximum discrepancy of about 0.6\%.\\ 
In Fig.~\ref{fig1}, we show the energy values as a function of the internuclear distance for the $^1\Sigma_g^+$, $^1\Sigma_u^+$, $^1\Pi_g$ and $^1\Pi_u$ symmetries obtained from our technique and compared with the results of Sánchez and Martín {\it et al} \cite{Martin}.
Note that there is an overall good agreement and, in the case of the second root of $^1\Sigma_u^+$ symmetry, our curve lies below that of Sánchez and Martín, which from the variational point of view is an indicator of the quality of our results. In Fig.~\ref{fig2}{\it a} note that the CI wavefunctions of the $^1\Sigma_g^+$ $Q_2$ in FC region are mainly composed of the $2p\pi_u;n\lambda\pi_u$ configurations, clearly converging to the $^2\Pi_u$ H$_2^+$ state. Figs ~\ref{fig1}{\it b}, \ref{fig1}{\it c} and \ref{fig1}{\it d} show some $Q_2$ states, namely, $^1\Sigma_u$, $^{1,3}\Pi_{g,u}$ symmetries running parallel to the ionic hydrogen states (as also observed by Sánchez and Martín \cite{Martin}).\\

\begin{table}[htb!]
\centering
\caption{Dipole Transition Moment for some internuclear distances calculated by (a) Borges \cite{Itamar}, (b) Wolniewicz \cite{Wolniewicz} and (c) This work}
\vspace{0.5cm}
\begin{tabular}{c c c c c c c}
\hline
\hline
State&R=1.0$a_0$&R=1.2$a_0$&R=1.4$a_0$&R=1.6$a_0$&R=1.8$a_0$&R=2.0$a_0$\\
\hline
$B^1\Sigma^+_u$&$0.7269^{a}$&$0.8359^{a}$&$0.9236^{a}$&$1.025^{a}$&$1.125^{a}$&$1.218^{a}$\\      
				&$0.7650^{b}$&$0.8708^{b}$&$0.9821^{b}$&$1.096^{b}$&$1.208^{b}$&$1.313^{b}$\\
				&$0.7654^{c}$&$0.8707^{c}$&$0.9819^{c}$&$1.095^{c}$&$1.207^{c}$&$1.313^{c}$\\  
$B'^1\Sigma^+_u$&$0.3262^{a}$&$0.3584^{a}$&$0.3886^{a}$&$0.4148^{a}$&$0.4352^{a}$&$0.4478^{a}$\\
				&	   &	  &$0.3966^{b}$&	    &$0.4355^{b}$&$0.4388^{b}$\\
				&$0.3394^{c}$&$0.3709^{c}$&$0.3990^{c}$&$0.4210^{c}$&$0.4341^{c}$&$0.4366^{c}$\\
\hline
\hline
\end{tabular}
\label{table1}
\end{table}

\subsection{Intermediate Range}
\begin{figure}[htbp]
\centering
\includegraphics[width=6.3in,height=6.3in,keepaspectratio]{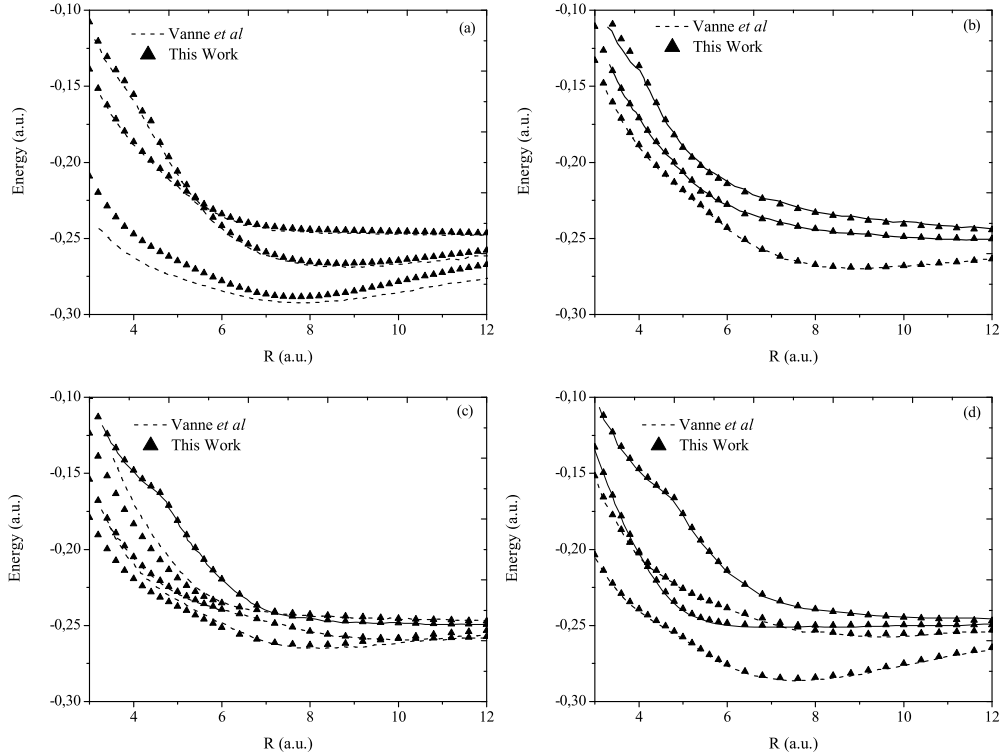}
\caption{Potential Energy Curves as a function of the internuclear distance. Dashed lines for Vanne {\it et al} \cite{Dalgarno} curves and triangles for this work: (a) $Q_2$ $^1\Sigma_g^+$ symmetries,(b) $Q_2$ $^3\Sigma_u^+$ symmetries, (c) $Q_2$ $^1\Pi_u$ and $^3\Pi_g$ symmetries and (d) $Q_2$ $^1\Pi_g$ and $^3\Pi_u$ symmetries}
\label{fig3}
\end{figure}

\begin{table}[htb!]
\centering
\caption{Some values of electronic energy for the $Q_2$ states (a) This work, (b) Vanne {\it et al} \cite{Dalgarno} and (c) Sánchez and Martín \cite{Martin}}
\vspace{0.5cm}
\begin{footnotesize}
\begin{tabular}{p{0.8cm}p{1.5cm}p{1.5cm}p{1.5cm}p{1.5cm}p{0.8cm}p{1.5cm}p{1.5cm}p{1.5cm}p{1.5cm}}
\hline
\hline
State&R=3.0$a_0$&R=4.0$a_0$&R=5.0$a_0$&R=6.0$a_0$&State&R=3.0$a_0$&R=4.0$a_0$&R=5.0$a_0$&R=6.0$a_0$\\
\hline
$1^1\Sigma^+_g$&$-0.20900^a$&$-0.24715^a$&$-0.26463^a$&$-0.27781^a$&$1^3\Sigma^+_u$&$-0.13314^a$&$-0.18855^a$&$-0.21811^a$&$-0.24277^a$\\
			   &$-0.21167^b$&$-0.24969^b$&$-0.26679^b$&$-0.27929^b$&               &$-0.13404^b$&$-0.18940^b$&$-0.21896^b$&$-0.24371^b$\\
		       &$-0.20694^c$&$-0.24595^c$&$-0.26350^c$&$-0.27491^c$&               &$-0.13201^c$&$-0.18717^c$&$-0.21584^c$&$-0.23600^c$\\  
			   &		&	 	 &	    	&\\
$2 ^1\Sigma^+_g$&$-0.13883^a$&$-0.18658^a$&$-0.21412^a$&$-0.24170^a$&$2^3\Sigma^+_u$&$-0.11065^a$&$-0.17072^a$&$-0.20604^a$&$-0.22782^a$\\
		&$-0.13888^b$&$-0.18686^b$&$-0.21508^b$&$-0.24338^b$&                       &$-0.11086^b$&$-0.17110^b$&$-0.20702^b$&$-0.22860^b$\\
		&$-0.13618^c$&$-0.18469^c$&$-0.21250^c$&$-0.23579^c$&                       &$-0.10778^c$&$-0.16762^c$&$-0.20193^c$&$-0.22502^c$\\
		&	&	&	&\\

$3 ^1\Sigma^+_g$&$-0.10606^a$&$-0.15542^a$&$-0.20582^a$&$-0.23386^a$&$3^3\Sigma^+_u$&$-0.08383^a$&$-0.13517^a$&$-0.19006^a$&$-0.21393^a$\\
		&$-0.10958^b$&$-0.15715^b$&$-0.20679^b$&$-0.23536^b$&                       &$-0.09039^b$&$-0.14035^b$&$-0.18971^b$&$-0.21329^b$\\
		&$-0.10816^c$&$-0.15465^c$&$-0.20082^c$&$-0.23150^c$&                       &$-0.08971^c$&$-0.13960^c$&$-0.18343^c$&$-0.20770^c$\\
		&	&	&	&\\

$1 ^1\Pi_u$	&$-0.17898^a$&$-0.21929^a$&$-0.23739^a$&$-0.25124^a$&$1^3\Pi_u$     	&$-0.20372^a$&$-0.23918^a$&$-0.25745^a$&$-0.27550^a$\\
		&$-0.18051^b$&$-0.22099^b$&$-0.23896^b$&$-0.25327^b$&						&$-0.20540^b$&$-0.24084^b$&$-0.25885^b$&$-0.27663^b$\\
		&$-0.18817^c$&$-0.21884^c$&$-0.23629^c$&$-0.24904^c$&						&$-0.20225^c$&$-0.23739^c$&$-0.25323^c$&$-0.26958^c$\\
		&	&	&	&\\

$2 ^1\Pi_u$	&$-0.12364^a$&$-0.18319^a$&$-0.21884^a$&$-0.23519^a$&$2^3\Pi_u$			&$-0.13281^a$&$-0.20133^a$&$-0.23889^a$&$-0.24837^a$\\
		&$-0.12739^b$&$-0.18740^b$&$-0.22278^b$&$-0.23782^b$&						&$-0.13397^b$&$-0.20239^b$&$-0.23994^b$&$-0.24944^b$\\
		&$-0.12497^c$&$-0.18456^c$&$-0.21922^c$&$-0.23442^c$&						&$-0.13171^c$&$-0.19904^c$&$-0.23766^c$&$-0.24772^c$\\
		&	&	&	&\\

$1 ^1\Delta_g$	&$-0.21188^a$&$-0.24838^a$&$-0.26025^a$&$-0.26234^a$&$1^3\Delta_u$	&$-0.12873^a$&$-0.18491^a$&$-0.21369^a$&$-0.22964^a$\\
		&$-0.21453^b$&$-0.25106^b$&$-0.26306^b$&$-0.26553^b$&						&$-0.12966^b$&$-0.18595^b$&$-0.21468^b$&$-0.23054^b$\\
		&$-0.21060^c$&$-0.24768^c$&$-0.26044^c$&$-0.26332^c$&						&$-0.12851^c$&$-0.18470^c$&$-0.21349^c$&$-0.22936^c$\\
\hline
\hline

\end{tabular}
\end{footnotesize}
\label{tabela4}
\end{table}
The range $3 a.u.\le R\le 12 a.u.$ can still be considered as intermediate regime, in the sense that in this region the potential energy wells of the $Q_2$ states can still be found. In this region, there is still electronic correlation, however smaller. This region is important for the verification of crossings between states, which are of fundamental importance in calculations of survival probability in a dissociative regime. The states were calculated firstly with a careful choice of the active space, including the $2s\sigma_g$, $2s\sigma_u$, $2p\pi_u$, $2p\pi_g$, $3p\pi_u$, $3p\pi_g$, $3s\sigma_g$ and $3s\sigma_u$ orbitals. As mentioned before, the doubly excited states are characterized by the exclusion of configurations with occupation on the orbitals of type $1\sigma_g$ $(Q_1)$ or $1\sigma_g$ and $1\sigma_u$ ($Q_2$). In the Fig.~\ref{fig3}, we present the potential energy curves for the $^{1,3}\Sigma^+_g$, $^{1,3}\Sigma^+_u$, $^{1,3}\Pi_u$ and $^{1,3}\Pi_g$ states compared with the data of Vanne {\it et al} \cite{Dalgarno} at the $3.0 a.u.\le R\le 12 a.u.$ range and, as can be seen, our results show a discrepancy smaller than 1\%. Here we should emphasize that our basis-set is in a number of 110 molecular orbitals, whereas Vanne's calculation took about 150 functions. Values of electronic energy of the singlets and triplets for some internuclear distances also are presented in Table \ref{tabela4}. 
It must be noted that the advantage of our technique is that we are able to work in all internuclear regions without concerning about the adequacy of the basis-set to a certain internuclear region, which is not possible in the case of Hartree-Fock based CI approach \cite{Martin}. Another advantage over CI with B-spline functions methods, as we will see on the next section, is that by using a  relaxed basis we are able to uniquely identify the way each state dissociates, even in the degenerate cases. 
\subsection{Asymptotic Range}
\begin{figure}[h!]
\begin{minipage}{0.45\linewidth}
\centering
\includegraphics[width=\textwidth]{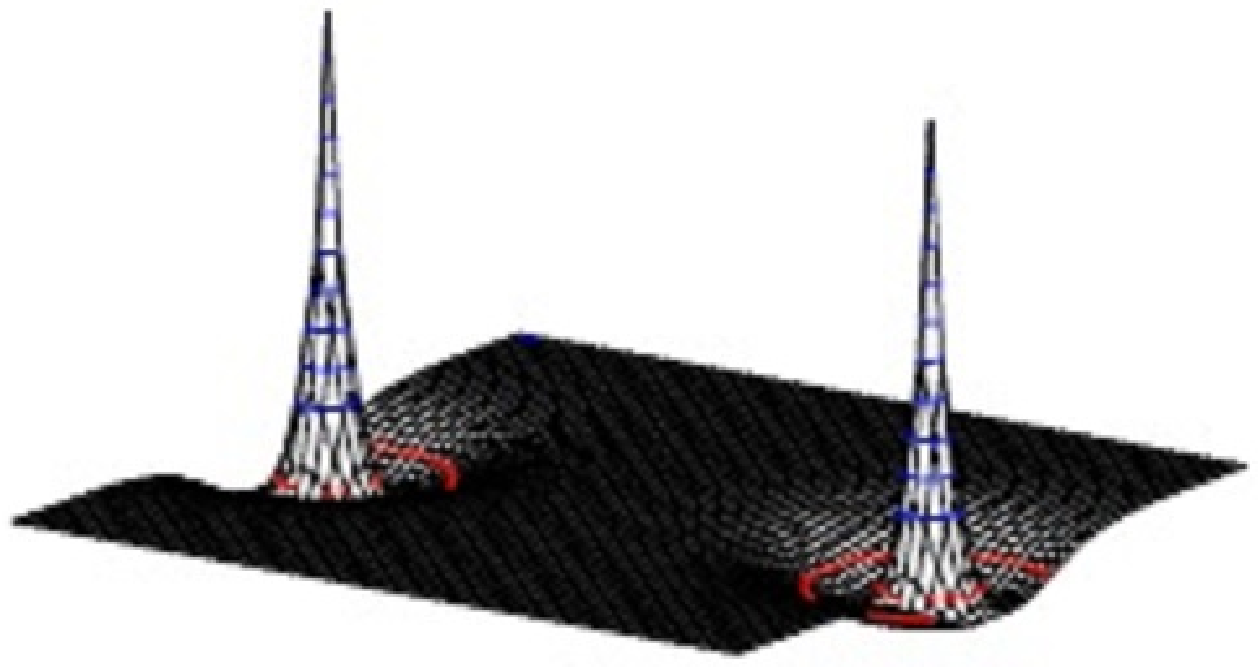}
\end{minipage}
\begin{minipage}{0.45\linewidth}
\centering
\includegraphics[width=\textwidth]{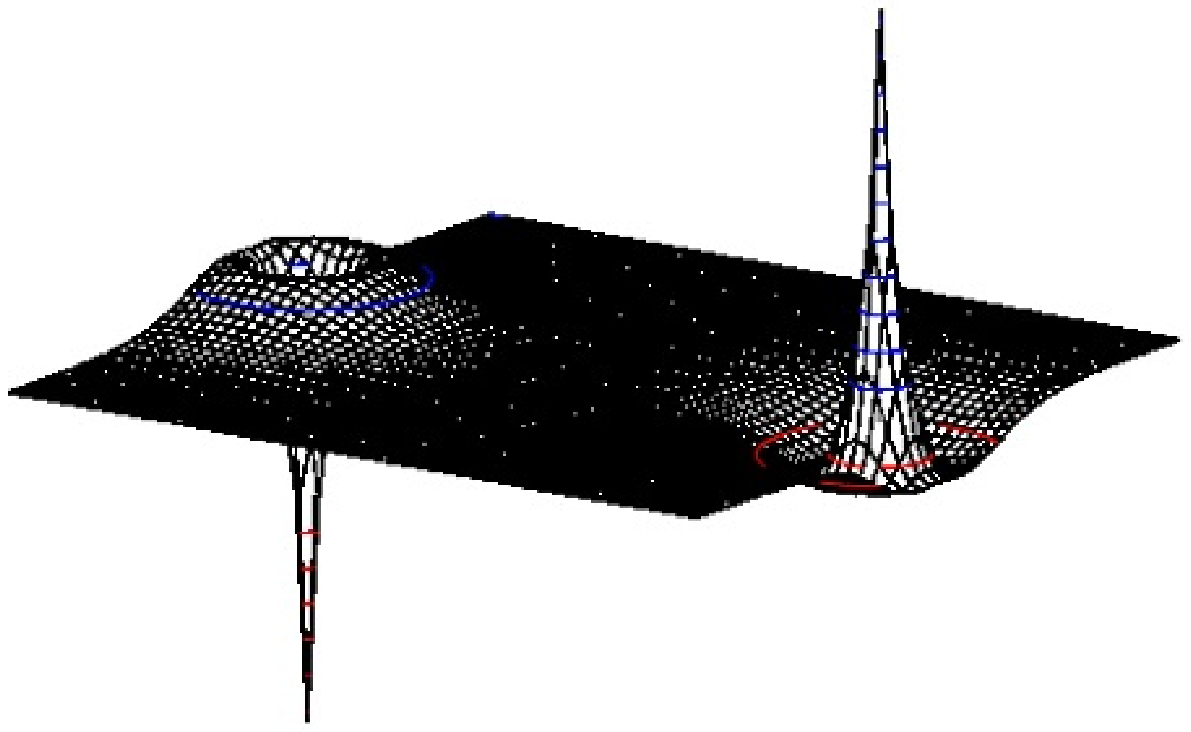}
\end{minipage}
\caption{In the left, 3D electronic amplitude plot of $2s\sigma_g$; in the right, the same for $2s\sigma_u$ orbital. Colour scheme: blue line stands for a positive amplitude value and red line for negative amplitude (online version)}
\label{fig4}
\end{figure}

In this internuclear distance, the electronic correlation is almost zero and only a few dispersion terms of the order of $1/R^6$ may play a role between the atoms via van der Waals interaction \cite{Vanderwaals}. As said before, one of the fundamental differences between the MCSCF and HF-based CI methods is that in the latter, the orbitals remain fixed while only expansion coefficients of the configurations are taking as variational parameters. However, in MCSCF the orbitals and CI coefficients are both optimized. This relaxation of the orbitals allows us, through the analysis of the configuration coefficients of higher weight, to determine whether the molecule is going or not to dissociate on the H(2s) + H(2s) fragments.
The main configuration of the $1Q_2$$^1\Sigma_g^+$, whose dissociation energy is -0.25 a.u. (two hydrogen atoms with $n=2$) in the asymptotic region, is composed by $2s\sigma_g$ and $2s\sigma_u$ molecular orbitals. The shape of these molecular orbitals is shown in Fig.~\ref{fig4} at R=80 a.u., in which the electronic amplitudes are plotted. 3D plots for the electronic amplitude are shown for both the $2s\sigma_g$ (on the left) and $2s\sigma_u$ (on the right) orbitals. It is evident from these pictures that, at the separated atoms regime, both atomic orbitals present one node which is characteristic of a 2s orbital. 

It is important to note that neither the work of Sánchez and Martín \cite{Martin2} nor any other theoretical work on doubly excited states of H$_2$ molecule were able to provide information about the $Q_2$ electronic states in the dissociation limit, given that the states with $\Sigma_{g,u}^+$ symmetry may dissociate in H$(2l)$+H$(2l’)$, where l and l’ are either s or p, that is, they are degenerated and that requires, from the theoretical perspective, more information than simply the energy to be able to determine which fragments were produced.

\subsection{Generalized Oscillator Strength}

\begin{figure}[htbp]
\centering
\includegraphics[width=6.3in,height=6.3in,keepaspectratio]{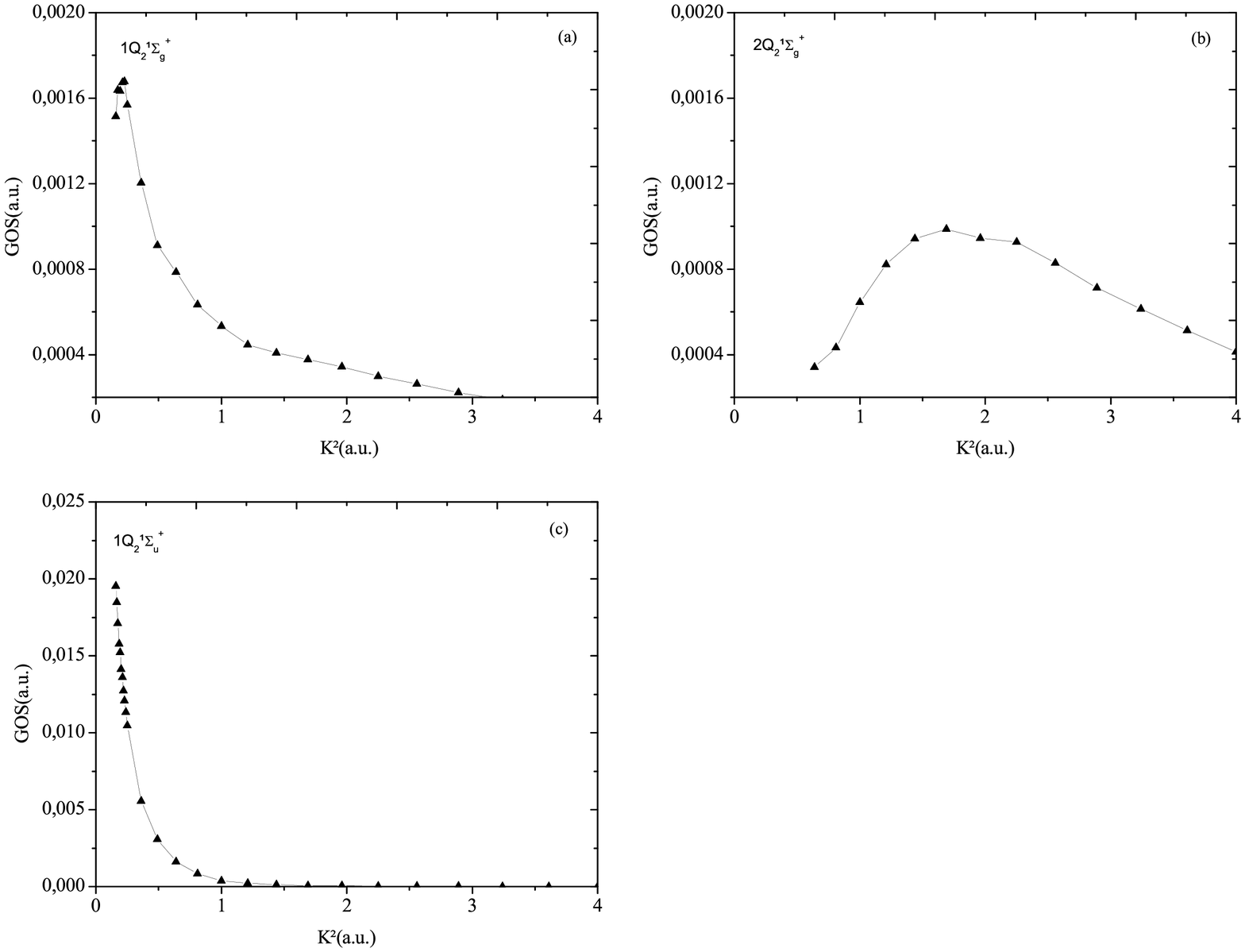}
\caption{Generalized Oscillator Strength as a function of the transferred momentum $K^2$ distance for doubly excited states within MRCI method: (a) $1Q_2$ $^1\Sigma_g^+$ symmetry,  (b) $2Q_2$ $^1\Sigma_g^+$ symmetry and (c) $1Q_2$ $^1\Sigma_u^+$ symmetry}
\label{fig6}
\end{figure}

In Fig.~\ref{fig6} is presented the generalized oscillator strength as function of the transferred momentum for three doubly excited states of lowest energies ($1Q_2$ $^1\Sigma_g^+$, $2Q_2$ $^1\Sigma_g^+$ and $1Q_2$ $^1\Sigma_u^+$) of the H$_2$ molecule. MRCI wavefunctions have been used and the calculations have been done at the distance of equilibrium (1.4 a.u.) of the molecule. Note that, there are no previous results to be compared. In the Fig.~\ref{fig6}{\it a}, for the $1Q_2$ $^1\Sigma_g^+$, it can be seen that the GOS curve presents a maximum peak at $K^2=0.4 a.u.$. This is a very useful information that can be used to set an optimum experimental apparatus since the GOS is directly related to the excitation cross-section. For the same reason of the early state, Fig.~\ref{fig6}{\it b} shows that the transition for the second $^1\Sigma_g^+$ is forbidden, presenting a smoother peak at $K^2=1.8 a.u.$. Finally, for the state $^1\Sigma_u^+$ the transition is allowed. The corresponding profile is shown in Fig. 5c. 

\section{Conclusions}

One of our goals, namely, to map the $Q_2$ state which dissociates in H(2s) + H(2s) was achieved successfully. Here, again, we should emphasize that no other technique was capable to provide this crucial information in what concerns the applications in twin atoms \cite{Grupo1}. Further, we are able to plot energy curves up to a higher internuclear distance than that of Sánchez and Martín \cite{Martin2}, given that their B-spline functions present convergence problems for distances greater than 6 a.u.. Although our base is composed by Gaussian functions centered at the nuclei, it was still possible to reach convergence at distances higher than 100 a.u.. We have also performed calculations for the transition moments for the states $B^1\Sigma_u^+$ and $B'^1\Sigma_u^+$ and achieved precise results and agreement comparing to previous calculations and the exact value \cite{Wolniewicz}. We also were able to perform matrix elements calculations in the FBA for some doubly excited states and here it is worth pointing out that, as said before, a transition calculation for a doubly excited state in the FBA (which is an one-electron operator), should be identically null according to Condon-Slater rule, for a single determinant approach but, our highly correlated wavefunction allowed us (despite of an one-electron operator calculation) to be able to lead to a set of non-null values for the transition matrix. This gives us a solid and promising basis for the future cross section calculations as well as other physical quantities of interest from the theoretical and experimental point of view.
We intend in the future to refine this approach, in order to extract more information about the excitation
cross-section.

\ack

This work is supported by CAPES, FAPERJ and CNPq.

\end{document}